# DOUBLE BOUNDARY and PARITY DOUBLING in HADRONS

A. E. Inopin

Karazin National University, Department of Experimental Nuclear Physics,
Svobody Square 4, 61077, Kharkov, Ukraine

**Abstract.** A new mechanism for the parity doublers in hadrons is suggested.

## I. Introduction, and View "Through the Looking Glass"

Black holes (BH) are one of the most intriguing objects in nature, both theoretically and experimentally. Being predicted in 1916 by Karl Schwarzschild, until recently there has been no clear experimental evidence of the back holes [1].

Black holes are akin to the quarks – the major building blocks of matter. Although Zweig had predicted quarks in 1964, nobody has seen them materialize so far. Even more interesting, according to the existing theory of quantum chromodynamics (QCD), colored quarks have to be *confined* to some finite radius (~1 Fermi), so all the observable hadrons are white.

In the case of black holes there is a total analogy *in reverse*–event horizons, located at $R_h = 2M$ ($M$ is the black hole's mass). If some material point, a human being, space craft or even light rays fall through the horizon, it will never again come outside of the black hole, and will be totally destroyed by the gravitational forces.

Now, the radius of the black hole strongly depends on its mass and could vary from the billions of solar masses $M_\odot$ to the scale of the proton's mass (1 GeV) and even less for the so-called elementary or quantum black hole. Here, on the proton's mass scale (PMS], these two seemingly unrelated phenomena merge. At this *quark-hadron scale* of a proton's mass and *higher*, chiral symmetry breaking is restored and the vacuum does not feel the apparent existence of quark and gluon condensates, which spoil the chiral symmetry from the start (the mechanism of spontaneous breaking of chiral symmetry, SBCS).

At the PMS there is a general mechanism, let's call it *Quark-Black Hole Confinement* (QBHC), which is responsible for the dynamics. It is based on the appearance of the *critical radius* of the 1 Fermi scale and the appropriate generalized dynamics – effective gravito-strong interaction (see Figure 1).

In gravity plus electromagnetism, there is another interesting mechanism possible – radiation trapping just on the horizon's surface – the formation of the so-called *horizon geon* (HG) [2]. We can show how the HG confinement picture (*confinement of light*) intertwines with quark's confinement mechanism, creating QBHC.

All together, this brings us to the concept of *Quark-Black Hole Duality* (QBHD). Quark-gluon and black hole bags are *dual* to each other. Quark-gluon confinement which traps matter from breaking inside out is dual to the black hole confinement which traps everything coming from outside to the black hole's mouth. This rich concept of duality will help in practice to compute observables in *time-like* regions, knowing the physics in *space-like* region and vice versa, which is in accord with M.C. Escher's duality [3].

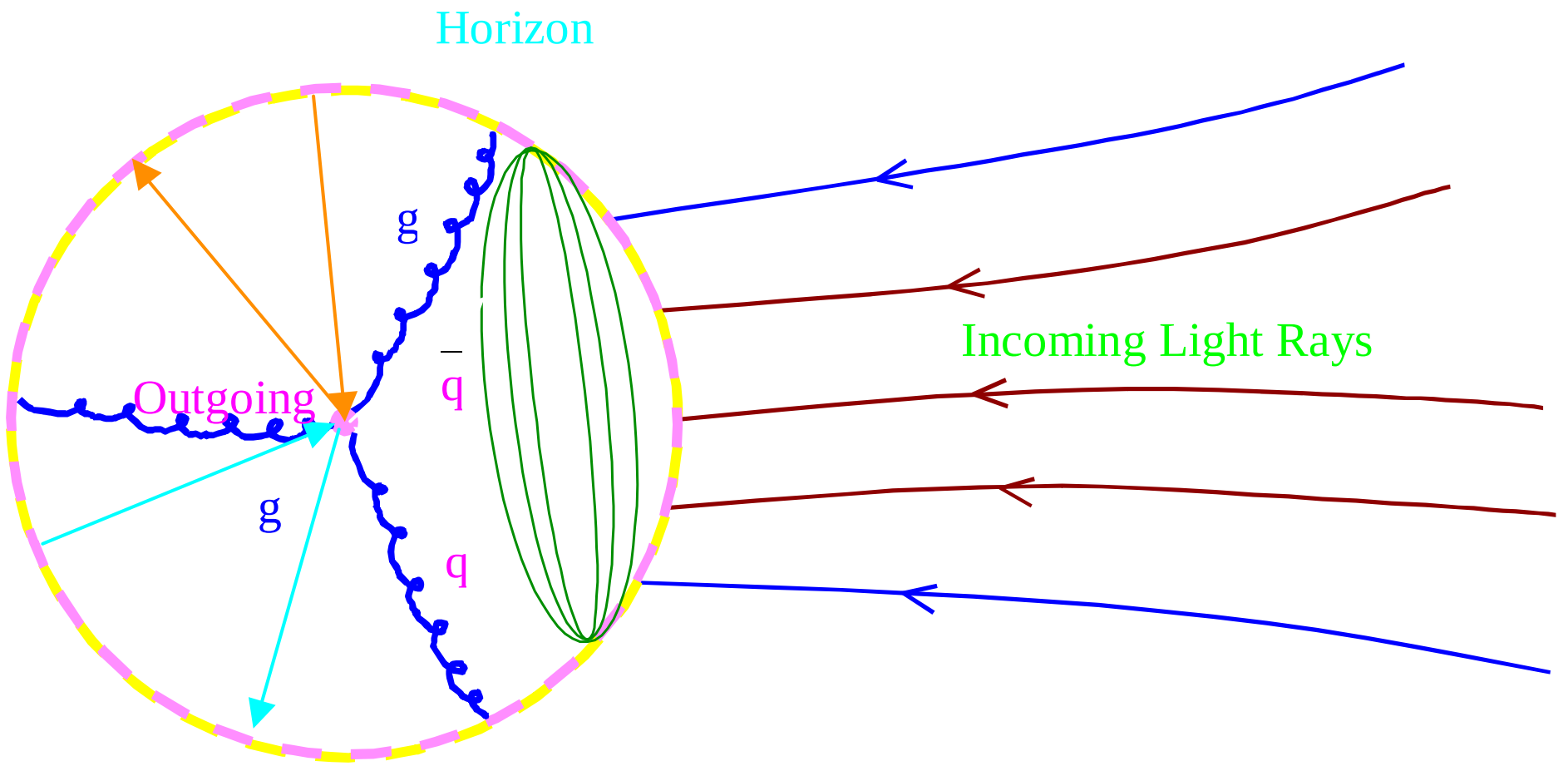

**Fig.1 (color online) Quark – Black Hole Bag**

As we know very well, quark-gluon bags could be not only of spherical shape but cylindrical as well. This is well matched by a quark-black hole's self-consistent dynamics, leading to the different forms of black holes (even a funnel shape is possible).

At this point let's return again to our quark–black hole bag. In considering these dual fields, the idea of *two scales* comes up naturally. We can divide all of the physics space into *two regions* – one is $0 < r < R_{\text{confinement}}$, and we term it *fast scale* (or short distance scale). The other is defined as $R_{\text{confinement}} < r < \infty$, and it is called *slow scale* (or large distances scale). In treating the dynamical problems, it is very convenient to separate the slow and fast degrees of freedom (Born-Oppenheimer approximation).

## II. Major Mechanisms

Now we are coming to the main point of this *Letter*. We are eager to dissect a real nature of the parity doubling phenomena in hadronic physics. Our *first* idea is that many hadrons have a double-confinement or double-lensing radius. This fact is almost evident from the results of DIS modeling of the hadronization process [4]. Quark-hadron duality in jet formation in DIS leads to two-step process of hadronization, with *two scales* appearing – large $Q_0^2 >> \Lambda^2$ and small $Q_0^2 \sim 1\text{GeV}^2$ ($\Lambda \equiv \Lambda_{QCD}$ everywhere in our paper). An alternative approach in DIS, termed as "Local Parton Hadron Duality" (LPHD) also leads to the *two dynamical scales*: $k_\perp = Q_0 \sim \Lambda$, and $k_\perp = Q_0 \sim 1$ GeV [4]. Both models of the hadronization process gives us the numbers vaguely in accord with our model - $R_1 \sim 0.2$-$0.3$ fm and $R_2 \sim 1$ fm.

Another fresh perspective can be taken with the so-called "Glue drops" model [5]. In this model the authors gave firm evidence of the existence of a non -perturbative scale, smaller than usual $1/\Lambda \sim 1$fm, which is related to gluonic degrees of freedom. The evidence for the presence of a *semi-hard* scale in hadronic structure is reviewed from many venues. The most notable effects are: QCD sum rules gives 0.3 fm for radius of the corresponding form factor, lattice gives 0.2-0.3 fm for the correlation length, energy of QCD string concentrated in a tube of radius 0.3 fm in transverse direction, instanton radius peaks approximately at 0.3 fm, diffractive gluon bremsstrahlung in hadronic collisions leads to $k_\perp$ for the gluons in proton of about 0.7 GeV. Observed slow rise of the total hadronic σ provides another evidence of small gluonic spots with transverse size $r_0 \sim 0.3$ fm [5]. The diffractive cone shrinks slowly with energy because of interactions with small drops inside proton, the same reason is for the weak gluon shadowing, and for the shortage of gluons at low scale. The valence quarks also get increased $p_\perp$, but only when $Q^2 >> 1/r_0^2$ in good agreement with data.

If there was only one scale in hadronic structure, the charge radius, one would expect the mean momenta of hadron constituents to be of the order of the inverse radius, i.e. $\langle k_\perp \rangle \sim \Lambda \approx 200$ MeV. This is what one would observe with a poor resolution, insufficient for seeing the structure of constituent quarks. However, with somewhat better resolution one can see gluons whose Fermi motion is much more intensive since they are confined within small spots, $\langle k_\perp \rangle \sim 1/r_0 \approx 700$ MeV. There should be manifestation of such an enhanced Fermi motion in reactions and observables sensitive to the primordial parton momentum.

It should be stressed that the double-scale naturally arise in a models describing restoration of the chiral symmetry breaking in hadrons [6-8]. The model explaining the parity doublers should interweave confinement and chiral symmetry breaking into *one single mechanism*, and this is exactly happens here. There are two different hadronic scales in the problem: $\Lambda_{QCD} \approx \Lambda_{BCS}$ and $\Lambda_{rest}$, and normally $\Lambda_{rest} > \Lambda_{QCD}$. We will return to the full discussion of double-radii systems in full-size article to be published elsewhere.

So, we claim that many mesons and baryons (including resonances) have this similar structure: it is a bag with first membrane being *semi-hard,* or semi-transparent, and second membrane is *hard*, impenetrable, and for these hadrons a parity doubling would occur. Quarks could also be thought of moving along the "caustics" inside a torus with a small internal radius.

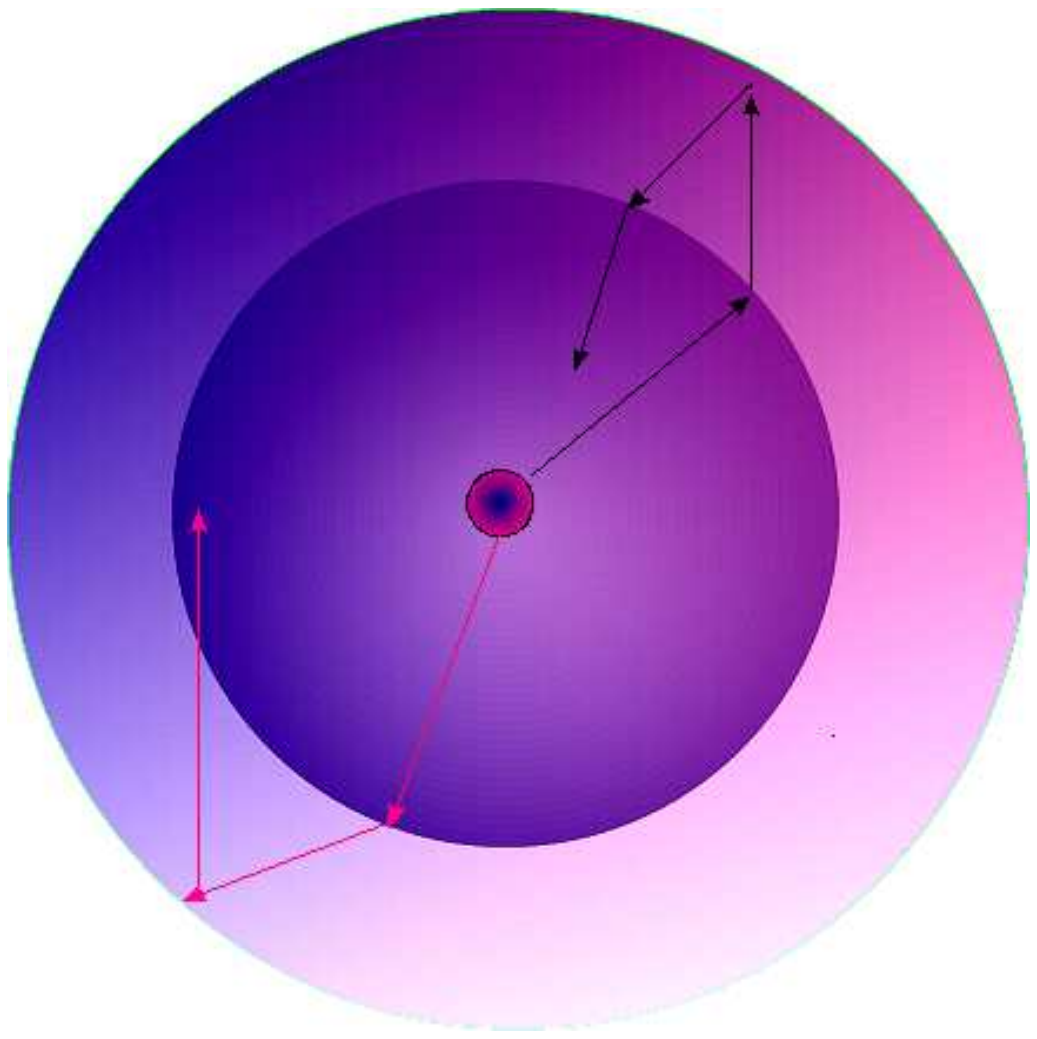

**Fig.2 (Color online) Dynamics inside the hadrons**

We can also draw a picture compatible with the following confinement scenario. From the Fig.1 one sees that the horizon is dual to the confinement boundary. This duality is performed via *lensing symmetry principle* [9]: quark's trajectories are trapped via horizon geon dynamics → they are "gliding" along the surface and are reflected back to the center. The boundary is acting as reflecting and focusing lens. Such a double boundary could have a different curvature – from almost planar hadron (high *J*), to the sphere, hyperboloid, etc: $q\bar{q}$ field lines are "closed" on the dual rays – as to be incoming to the black hole outside photons. Quarks are tracers and curled exactly on the quasi-shell $3M = R \equiv R_{\text{conft}}$. This is exactly the mechanism realized in black hole "knife-edge" orbit. Evidently, the hadron became a seashell, closed on R = 3M [10]. It will be a good idea to model a "hadronic lens" which will give such regimes. It is amazing to notice that Randall-Sundrum theory with extra dimensions [11] would apply to our dual mechanisms of parity doublers and quark-black hole bags.

Einstein rings gave birth to the proper topology. The formation of multiple images of a single star work in favor of the *double-ring* idea. Einstein donut and diamond necklace created a double-radius topology. We see the diamond necklace in the center of the Einstein donut on the cover of the book [12]. As one can witness inside diamond necklace we see precisely the double-ring structure.

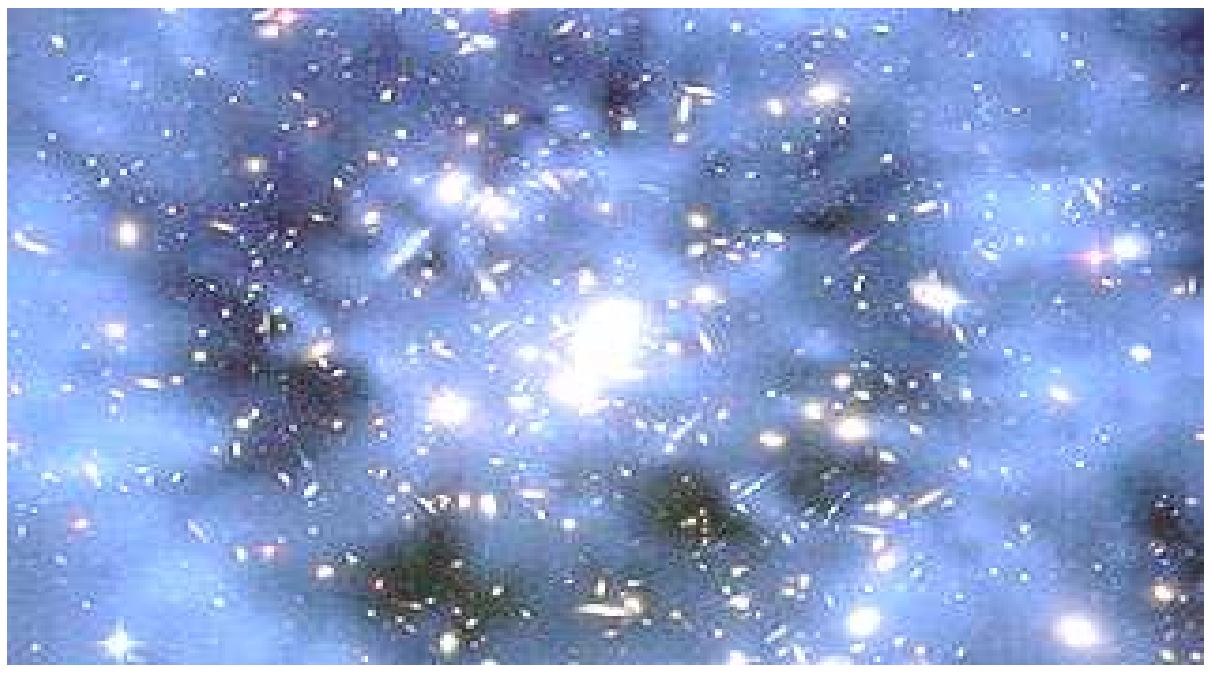

**Fig.3 (Color online) Ring of Dark Matter**

It is possible that the majority of hadrons can be *isomorphic* to the plane ***2D*** structures, in a sense that all their major properties are described by the spherical or elliptical 2***D***-projection and this very *degeneracy* gives birth to the parity doublers. So, parity doubling is *synonymous* of the term degeneracy, and Escher gave an example of how one can establish 2***D*** – 3***D*** correspondence [3]. We see here again the road to the t'Hooft and Maldacena holographic model for HEP -> all the ***3D*** properties could be inferred from the ***2D*** domain (or *ring*) [13]. It will be a good idea to connect a black hole's shells with $R$=2$M$ and $R$=3$M$ with order of magnitude for radiuses $R_1$ and $R_2$, which arises from DIS mechanism on $k_\perp$ [4].

And now, finally we are coming to our *second* idea. Quarks are flowing inside the hadron and from time to time they come close to the *first* wall – which is a semi-transparent membrane. They crossed this first membrane and reach for the *second* one – which is "black", perfect wall – *reflector*. So, there all quarks and gluons are reflected back to the center, and precisely the *P*-parity will be changed from "+" to "–" and vice versa. The Garcilazo's model exactly mimicked this property: $\Delta^+$ is *reflected* into $\Delta^-$, and $N^+$is *reflected* to $N^-$ via clusters through the zero point [14]. But we can also say, that this picture lead to a double boundary via reflections through the point zero. Baryons get together in islands and from there the exchanges proceed. Our parity doublers mechanism probably can be generalized on chiral quartets, which are seen in baryonic spectra as well [7].

One more explanation had crossed my mind. Parity doublers could be also rooted in dual space models. That could happen because two states are *spread* or scattered into the *different* spaces. Such a bifurcation can be measured by a huge global array of detectors, like Auger, HESS, or even LISA, which "sees" *both* spaces and could identify degenerate energies of both states.

One can see that the Polyakov-t'Hooft (HP) monopole with winding number *N*=1 is *topologically equivalent* to the torus structure with internal rays → which is exactly our double-radius hadron. The HP monopole with *N*=1 is also equivalent to our OZI-seashell [10]. The "ring structure" was just discovered by Hubble for dark matter and we will relate this ring to our double-radius hadrons in forthcoming paper (see Fig.3).

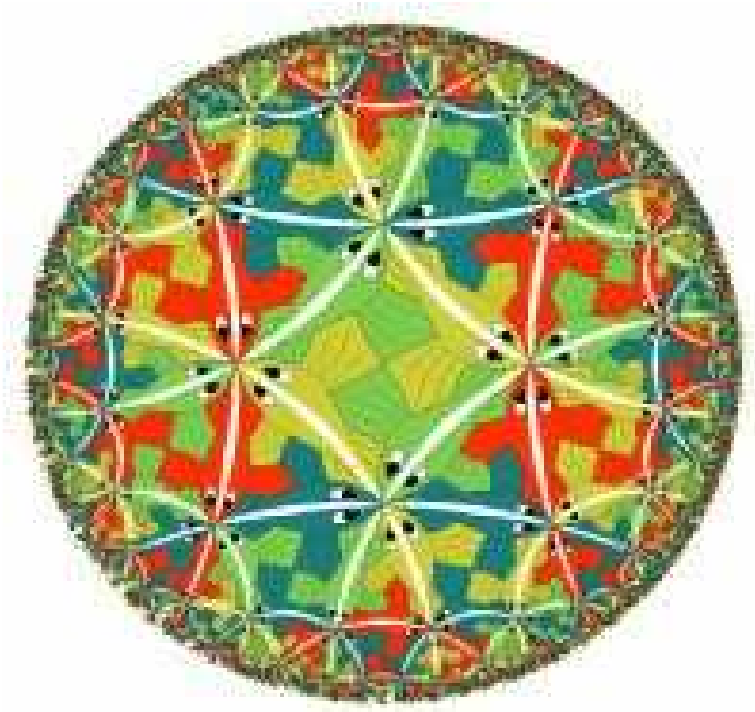

**Fig.4 Escher's Print "Circle Limit III"**

## REFERENCES

[1] V. Novikov and I.Frolov, *Black Hole Physics*, Kluwer Academic Publishers, 1998.
[2] R. Dumse, A.E. Inopin, private communication.
[3] B. Ernst, *The Magic Mirror of M.C. Escher*, Taschen, 2007.
[4] W. Melnitchouk et al., Phys.Reports v.406 (2005) 127-301.
[5] B.Z. Kopeliovich et al., hep-ph/0607337.
[6] J. E. F. T. Ribeiro et al., hep-ph/0507330 v1.
[7] A.E. Inopin, hep-ph/0404048, C.F. Diether III and A.E. Inopin, physics/0601110 .
[8] S. S. Afonin, arXiv: 0704.1639 v2.
[9] G.S. Melnikov, http://314159.ru/melnikov/melnikov6.htm.
[10] A.E. Inopin, hep-ph/0702257.
[11] L. Randall, *Warped Passages*, Harper Collins Publishers, 2005.
[12] E.F. Taylor, J.A. Wheeler, *Exploring Black Holes. Introduction to General Relativity.* Addison Wesley Longman, Inc, 2000.
[13] J. Maldacena, Scientific American, v.11 (2005), pp.57-63.
[14] H. Garcilazo et al., hep-ph/0610288.